\documentclass[9pt,twocolumn,twoside,prl,superscriptaddress]{revtex4}

\usepackage{braket}
\usepackage[normalem]{ulem}

\usepackage{multirow}
\usepackage{slashed} 
\usepackage{cancel}  
\usepackage{pdfpages}
\usepackage{amsmath,amsfonts}
\usepackage{graphicx}
\usepackage{dcolumn}
\usepackage{bm}
\usepackage{hyperref}
\usepackage{color}
\usepackage{xcolor}
\usepackage{mathtools}
\usepackage{float}
\usepackage{bm}
\usepackage{emptypage}
\usepackage{orcidlink}

\begin{document}

\title{Consistent determination of stability regimes in natural ecological communities from abundance time series}
 
\author{Alberto Meg{\'\i}as~\orcidlink{0000-0002-7889-1312}}

\affiliation{Complex Systems Group and Department of Applied Mathematics,
Universidad Polit\'ecnica de Madrid, Madrid, Spain}

\author{Vicente J. Ontiveros~\orcidlink{0000-0001-8477-2574}}

\affiliation{Ecology and Complexity Group, CEAB-CSIC, Blanes, Spain}

\author{David Alonso~\orcidlink{0000-0002-8888-1644}}

\affiliation{Ecology and Complexity Group, CEAB-CSIC, Blanes, Spain}

\author{Jos\'e A. Capit\'an~\orcidlink{0009-0007-3950-5988}}
\email{ja.capitan@upm.es}
\affiliation{Complex Systems Group and Department of Applied Mathematics,
Universidad Polit\'ecnica de Madrid, Madrid, Spain}

 
 
 
 
\keywords{Community stability $|$ Environmental variability $|$ Interaction inference $|$ Abundance time series $|$ Generalized Lotka--Volterra $|$ Mean-field theory }
 
\begin{abstract}
The stability of an ecological community is conventionally defined through species interactions, which quantify how species affect one another. Interaction strengths are notoriously difficult to measure in species-rich assemblages. Long-term monitoring of species-rich communities, however, provides species abundance time series, increasingly available across habitats and taxa but not yet connected to the stability properties of the communities they describe. Here we develop a statistical framework that infers stability regimes in large ecological communities directly from abundance time series. We study stochastic Generalized Lotka--Volterra dynamics with random interactions and environmental fluctuations using dynamical mean-field theory, reducing the multispecies system to an effective stochastic process for a representative species. The theory predicts three stability regimes (stable coexistence, intermittent dynamics close to extinction, and unbounded growth) separated by analytical boundaries, and shows that environmental stochasticity systematically destabilizes coexistence by promoting intermittent low-abundance dynamics. The resulting steady-state species abundance distribution is a Gamma law that ties the dynamical phases to the variability-based stability metrics used in empirical studies. Recasting the effective dynamics as a multivariate regression model, we infer interaction statistics, environmental variability and the characteristic timescale of the dynamics from community data, without reconstructing the interaction network. Applied to natural communities spanning a broad range of habitats and taxa, the method resolves contrasting stability regimes and yields quantitative estimates of species extinction risk.
\end{abstract}
 
 
\maketitle
 
 

From soil microbiomes to tropical forests, natural communities sustain large numbers of coexisting species, and the stability and persistence of that diversity underpins the functioning and maintenance of the ecosystems they form. Since the pioneering work of May~\cite{May1972}, ecologists have sought to explain how large numbers of interacting species can coexist despite environmental variability, demographic fluctuations, and the intrinsic complexity of ecological networks. Stability is a multifaceted concept~\cite{Grimm1997}, encompassing resistance to perturbations of equilibrium states~\cite{May1972,AllesinaTang2012}, resilience following disturbances~\cite{Holling1973,Pimm1984}, the permanence and persistence of species and communities~\cite{Law1996}, and the compatibility of environmental conditions allowing species coexistence, known as structural stability~\cite{Rohr2014,GrilliFeasibility2017,Saavedra2017,servan2018coexistence,lechon2026robust}. For communities that are permanently forced by a fluctuating environment, the equilibrium-based stability notions lose their operational meaning: there is no fixed point to return to, and stability is more naturally quantified by the magnitude of the fluctuations themselves. This is the variability view of stability~\cite{LoreauMazancourt2013,Wang2017,Arnoldi2016}, in which the stability of a population is measured by the inverse of its temporal coefficient of variation. 
 
Early random-matrix models suggested that increasing complexity destabilizes ecological systems~\cite{May1972}, whereas subsequent work demonstrated that stability can emerge from  interaction structures, including higher-order interactions, trophic organization, niche differentiation, and self-regulation~\cite{McCann2000,AllesinaTang2012,grilli:2017,lechon2026robust}. More recently, statistical physics approaches have shown that the collective behavior of diverse communities depends primarily on a small set of statistical properties of interaction matrices rather than on the detailed topology of ecological networks~\cite{Bunin2017,BarbierEtAl2018,grilli2020macroecological,CuiEtAl2023}. 

The interplay between both the intensity of species interactions and environmental variability is key to evaluate ecosystem stability. Thus, assessing the stability of natural communities requires connecting biotic interactions and environmental forcing with observational data through well-founded theory. Most ecological datasets consist of species abundance time series, whereas community models are formulated in terms of interaction coefficients, which are notoriously difficult to measure directly. Measuring interaction strengths directly has a long history, based on press or removal experiments~\cite{Paine1992}, on fitting pairwise coefficients from mono- and co-culture growth curves~\cite{FriedmanEtAl2017}, or on allometric and trait-based proxies~\cite{Berlow2004}. None of these strategies scales to the species-rich assemblages that long-term monitoring actually samples, and press-perturbation estimates recover elements of the inverse community matrix, so that indirect effects are unavoidably folded into the measurements~\cite{Novak2016}. Moreover, pairwise interaction measurements often fail to capture the indirect and collective effects that emerge in multispecies communities~\cite{Vandermeer1969,Vandermeer1981,ChangYu2023}. 
 
Statistical inference from observational data is the only route available at scale, as illustrated by the reconstruction of effective pairwise interactions among tropical tree species from static co-occurrence across census quadrats~\cite{VolkovEtAl2009}. Static co-occurrence and correlation patterns in abundances are, however, not identifiable with interactions: correlations may arise from shared environmental responses or phylogenetic relatedness, and very different networks can generate nearly indistinguishable correlation patterns~\cite{BlanchetEtAl2020,CamachoMateu2024}. 

Temporal data partially break this degeneracy, and several methods exploit it~\cite{IvesEtAl2003,SteinEtAl2013,UshioEtAl2018}. They aim, however, at reconstructing the full $S\times S$ matrix, so that the number of parameters grows as $S^2$ while the data grow only as $S\times T$, with $T$ the number of censuses: identifiability requires $T\gtrsim S$, so that the number of conditions is larger than the number of unknowns~\cite{skwara2023modelling}. The last requirement is rarely met in ecological monitoring, where richness often exceeds the number of censuses by an order of magnitude. Network-level inference from time series has therefore remained confined to small assemblages or to heavily regularized reconstructions~\cite{GibsonEtAl2021}, and few studies have attempted to extract dynamical, interaction-related quantities from the abundance series that biodiversity monitoring has accumulated~\cite{Dornelas2025}. Consequently, inferring effective interaction statistics directly from abundance data for species-rich communities remains a major unresolved challenge in community ecology~\cite{OvaskainenEtAl2017,WeissEtAl2016}.
 
Instead of attempting to reconstruct full interaction networks, one can alternatively infer the statistical properties of the interaction matrix that govern the community's macroscopic dynamics~\cite{CamachoMateu2024}. Here, we develop a statistical framework that predicts the stochastic dynamics, stability regimes and extinction risk of species-rich ecological communities directly from abundance time series. Building on Dynamical Mean-Field Theory (DMFT) applied to the Generalized Lotka--Volterra (GLV) model with environmental noise~\cite{galla2018dynamically,CuiEtAl2023}, we derive an effective stochastic differential equation (SDE) describing a representative species that captures the average dynamics of the community. In contrast to previous DMFT studies~\cite{galla2018dynamically,suweis2024generalized}, and following recent approaches describing microbial diversity through stochastic population dynamics~\cite{grilli2020macroecological,CamachoMateu2024}, our framework explicitly incorporates environmental stochasticity, allowing us to derive analytical predictions for abundance statistics and coexistence regimes under realistic environmental fluctuations. Although environmental variability and self-regulation have been shown to reproduce macroecological patterns in microbial communities~\cite{grilli2020macroecological}, we consider the combined effect of inter-specific interactions and environmental noise on coexistence~\cite{CamachoMateu2024}. 
 
\begin{figure*}[t!]
\centering
\includegraphics[width=\linewidth]{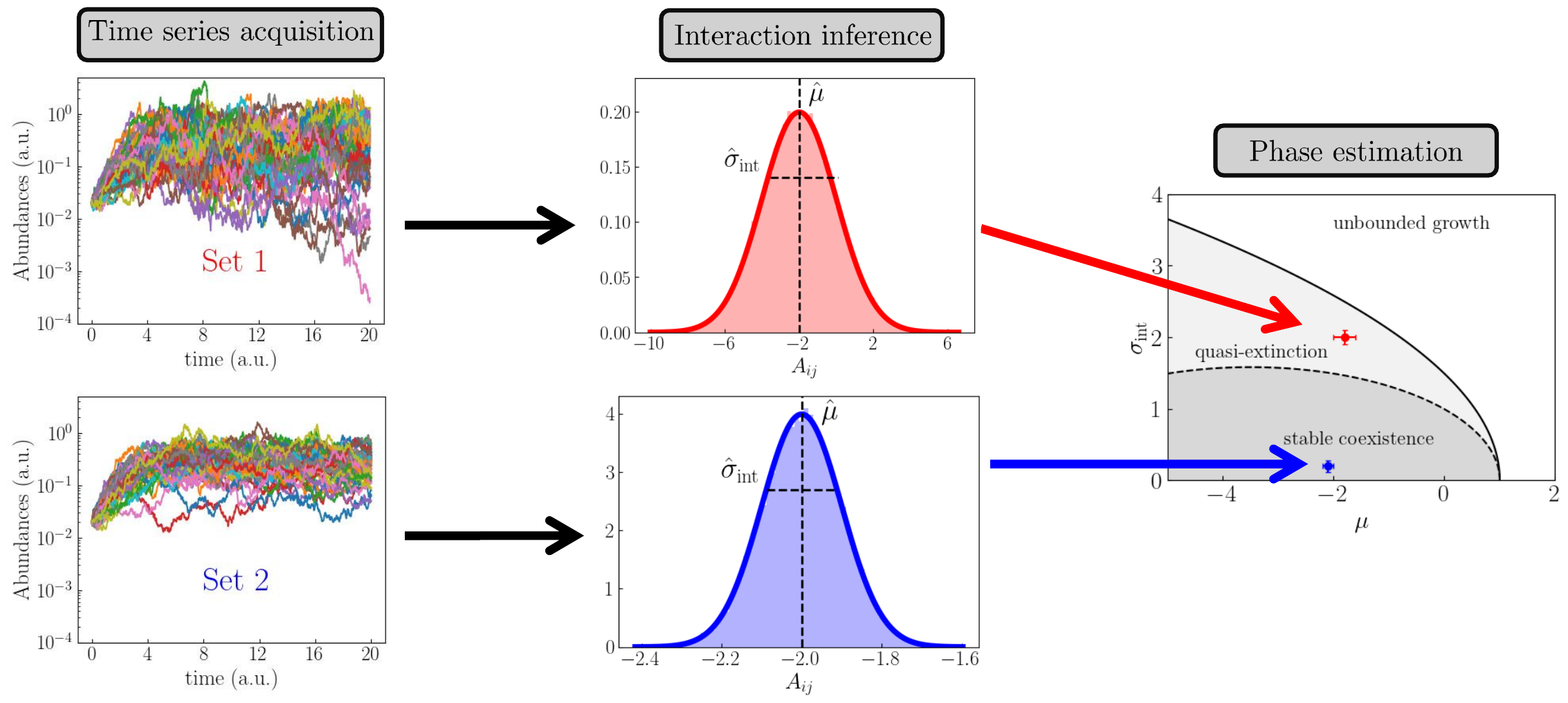}
\caption{\textbf{Statistical framework designed to infer stability regimes from abundance time series.} From a time series reporting species abundances in a natural community (left panels, two scenarios), the methodology allows to infer the first moments of the distribution of interactions (middle panels), as well as the intensity of the environmental noise. The inferred estimates inform about the community stability phase, this being either stable coexistence or quasi-extinction (right panel). Thus, species extinction risks in the community can be assessed quantitatively.}
\label{fig:conceptual}
\end{figure*} 
 
The SDE describing the effective species is recast as a multivariate regression model, enabling consistent inference of the first moments of the interaction distribution as well as the intensity of environmental noise from abundance time series (Fig.~\ref{fig:conceptual}). Estimated parameters are used to predict the dynamical regime of natural communities. The analysis of the corresponding Fokker--Planck equation yields a Gamma law for the species abundance distribution (SAD) at stationarity, consistent with empirical observations~\cite{WilsonEtAl2003,grilli2020macroecological,suweis2024generalized}. The shape of the SAD reveals three distinct stability regimes: a stable coexistence phase, a quasi-extinction regime where populations keep close to extinction, and an unbounded-growth region in parameter space, characteristic of GLV models~\cite{Bunin2017,galla2018dynamically,suweis2024generalized}.
 
The variability approach to stability~\cite{LoreauMazancourt2013,Wang2017,Arnoldi2016}, based on the coefficient of variation of temporal abundances, has the decisive practical advantage of being measurable directly from time series, without knowledge of interaction coefficients or of any underlying equilibrium. Our results connect these two approaches, yielding a quantitative threshold in variability that separates stable from unstable communities: the shape parameter $\alpha$ of the (Gamma) steady-state SAD predicted by the theory is equal to the inverse squared coefficient of variation of species abundance, so that the stable coexistence phase corresponds to variability (coefficient of variation) less than one, and the quasi-extinction phase is characterized by variability greater than one for the representative species time series. The phase boundary between stable coexistence and quasi-extinction phases is exactly the condition for the temporal standard deviation of the representative species' abundance equaling its mean abundance. Community stability regimes and species-level extinction risks are thereby expressed in the same framework that empirical studies of ecological variability already use. 
 
We validate our inference methodology using synthetic (simulation) data and apply it to long-term abundance records from a variety of natural communities, including tropical forests and plankton communities (Materials and Methods). Collectively, these empirical abundance datasets were selected because they are exceptionally well resolved in time or space, include systems with prior information on stability, and span a broad range of organisms~\cite{Vandeputte2021, Langenhaun2020, Pomati2012, Pomati2015, Ziolkowski2025, Ernest332783, Condit2019, Brown2005}. Across these systems, the method reconstructs interaction statistics from observational data and predicts their expected stability regime. By quantifying the probability that a species goes extinct in the quasi-extinction phase, our approach allows estimating extinction risks from biodiversity time series, and establishes a direct link between ecological observations, interaction statistics, and the dynamical stability of complex natural communities.
 
\section*{Results}
 
\subsection*{Community dynamics with environmental noise}
 
We consider the effect of environmental variability in a community of $S$ interacting species. The temporal variation of species abundances $x_i(t)$ is driven by stochastic GLV equations,
\begin{equation}\label{eq:SDE_GLV}
\frac{\dot x_i(t)}{x_i(t)}=r_i - x_i(t)+ \sum_{j\ne i}a_{ij}(t)x_j(t) +
\sqrt{2\sigma_{\mathrm{env}}^2}\,\xi_i(t),
\end{equation}
where $r_i$ are the intrinsic growth rates and $a_{ij}$ account for pairwise interactions. Environmental variability enters as a multiplicative, uncorrelated Gaussian white noise $\xi_i(t)$ added to the per-capita growth rate. $\sigma_{\mathrm{env}}^2$ controls the intensity of environmental fluctuations. Intraspecific interactions are equal to $-1$, which amounts to a rescaling of abundances (Materials and Methods). The  timescale of variation of interactions is assumed to be comparable to that of the environmental noise. Therefore, inter-specific interactions $a_{ij}$ ($i\ne j$) are modeled as random variables with mean $\mu/S$ and variance $\sigma_{\mathrm{int}}^2/S$,
\begin{equation}\label{eq:alpha}
a_{ij}(t) = \frac{\mu}{S} + \frac{\sigma_{\mathrm{int}}}{\sqrt S}z_{ij}(t),
\end{equation}
where $z_{ij}(t)$ are uncorrelated Gaussian random variables with mean zero and are also uncorrelated with environmental noise variables, $\xi_i(t)$. Growth rates are assumed to be normally distributed with an average $r$ and a certain variance, which can be subsumed, without loss of generality, into the environmental noise term. We provide a precise mathematical definition of the model in the Materials and Methods section.
 
Dynamical mean-field theory (DMFT) allows to collapse, in the limit of high diversity ($S\to\infty$), the multi-species population dynamics into an effective dynamics for a single, ``representative'' species. In the SI Appendix (section~S4, Eq.~S27) we show that the steady-state behavior of the effective, single-species dynamics remains unchanged if a non-zero correlation $\rho$ between off-diagonal interactions $a_{ij}$ and $a_{ji}$ is considered. 
 
For $\rho=0$, the effective dynamics reduces to the stochastic differential (SDE) equation,
\begin{equation}\label{eq:DMFT}
\frac{\dot x(t)}{x(t)}=r-\sigma_{\mathrm{env}}^2 -
x(t) + \mu M(t) + \sigma_{\mathrm{int}}\eta(t) +\sqrt{2\sigma_{\mathrm{env}}^2}\,\xi(t),
\end{equation}
where $M(t)=\langle x(t)\rangle$ is the average abundance of the effective species, where $\langle\cdot\rangle$ denotes an average over realizations of the stochastic process defined by the SDE in Eq~(\ref{eq:DMFT}), and $\eta(t)$ is a zero-mean Gaussian process whose second moment is equal to $\sigma_{\text{int}}^2\Sigma(t)$, where $\Sigma(t) := \langle x^2(t)\rangle$ is also determined self-consistently by abundance fluctuations. Therefore, the SDE has to be solved self-consistently to satisfy these two restrictions. The effective equation captures the collective behavior of the entire ecological community through self-consistent stochastic forcing. The SI Appendix, section~S2, details the DMFT derivation of the SDE for the representative species for arbitrary correlation $\rho$.
 
\subsection*{Inference of interaction statistics from abundance time series}
 
The effective stochastic dynamics enables direct statistical inference of interaction properties from abundance data. Discretizing Eq.~\ref{eq:DMFT} over a time interval $\Delta t$ yields a multivariate linear regression model for the logarithmic variation of the abundance of the representative species,
\begin{equation}
\log x(t+\Delta t)-\log x(t) = \beta_0 +\beta_1 x(t)+\beta_2 M(t) + e(t).
\label{eq:RM}
\end{equation}
At time $t$, $e(t):=\sqrt{\Delta t\,( 2\sigma_{\text{env}}^2+\sigma_{\text{int}}^2\Sigma(t))}\,\omega(t)$, where $\omega(t)$ is a standard normal random variable (uncorrelated in time by model assumptions), and regression coefficients depend on model parameters as $\beta_0:=\Delta t \,(r - \sigma_{\text{env}}^2)$, $\beta_1:=-\Delta t$, and $\beta_2:=\mu\,\Delta t$. We estimate $M(t)=\langle x(t)\rangle$ from temporal data as the average abundance $\widehat{M}(t)=\frac{1}{S}\sum_{i=1}^S x_i(t)$ at each time. 
 
Eq.~\ref{eq:RM} motivates a two-stage regression procedure (SI Appendix, section~S3, Fig.~S1). Because $\beta_i$ are time-independent, we can pool data across sampling times to improve statistical inference (this is specially important for short time series). Let $T$ be the length of the time series. We have $T-1$ observations of the logarithmic difference, $y_i(t):=\log x_i(t+\Delta t)-\log x_i(t)$, for $i=1,\dots S$ species. By pooling different times across the series, we have $n=S\times (T-1)$ observations for the response variable $y$, as well as for the regressors $x$ and $M$---for the latter, the average abundance $\widehat{M}(t)$ at fixed time is repeated $S$ times across species. The intercept and slopes of the linear model yield bootstrap estimates for $\Delta t$, $\mu$, and $r - \sigma_{\text{env}}^2$. 
 
In the second stage, we use the fact that the residual variance $\text{Var}[e_i(t)]=( 2\sigma_{\text{env}}^2+\sigma_{\text{int}}^2\Sigma(t))\Delta t$ depends linearly on the second moment $\Sigma(t)$, so $\text{Var}[e_i(t)]$ is regressed against abundance fluctuations to infer environmental and interaction variance parameters. We estimate $\text{Var}[e_i(t)]$ from data as $\frac{1}{S}\sum_{i=1}^S e^2_i(t)$, and the second moment of effective abundance, $\Sigma(t)$, by the sample second moment at fixed time, $\widehat{\Sigma}(t)=\frac{1}{S}\sum_{i=1}^S x^2_i(t)$. The parameters of the second linear regression model (i.e., the intercept $\sigma_{\text{env}}^2$, and slope $\sigma_{\text{int}}^2$, given that $\Delta t$ is already known) are inferred from $T$ observations through bootstrap replicates. Then the estimate $\beta_0$ of the intercept of the multivariate regression model, together with the estimate for $\sigma_{\text{env}}^2$, 
allow inferring the average growth rate $r$, and model parameters are completely identifiable from data.
 
\begin{figure}[t!]
\centering
\includegraphics[width=0.8\columnwidth]{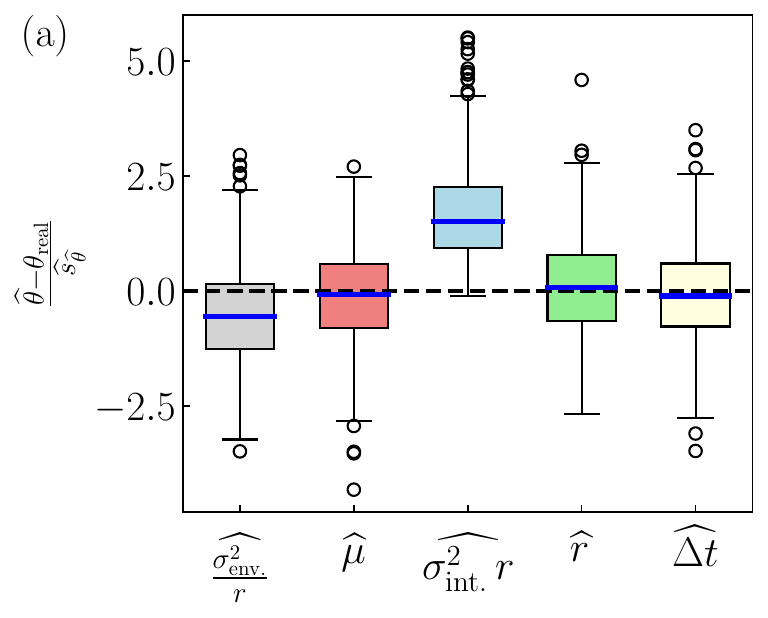}
\includegraphics[width=0.8\columnwidth]{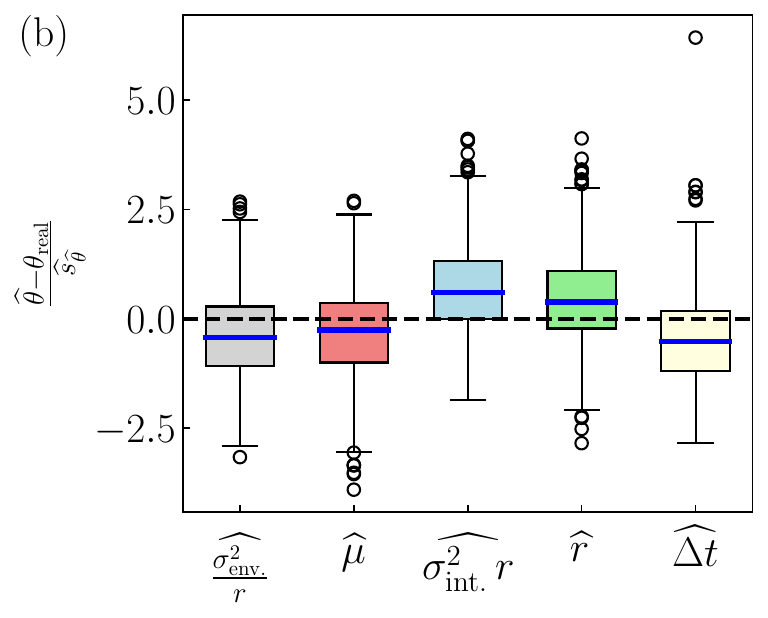}
\caption{\textbf{Validation of parameter estimates with GLV simulated dynamics.} The distributions of the $z$-score for each estimate are shown. Panel (a) shows estimation results in the absence of interactions: $\sigma_{\rm env}^2=0.1$, $\mu=\sigma_{\rm int}=0$. Panel (b): $\sigma_{\rm env}^2=0.05$, $\mu= -2$, $\sigma^2_{\rm int}=0.25$. Remaining parameters are: $r=1$, $\Delta t=0.02$, $S=50$ interacting species. Both variances, $\sigma_{\rm env}^2$ and $\sigma_{\rm int}^2$, are re-scaled with $r$, see Eq.~\ref{eq:UG} and SI Appendix, section~S4.2.}
\label{fig:validation}
\end{figure} 
 
Tests using synthetic stochastic simulations show accurate recovery of the true parameter values used in simulations. Fig.~\ref{fig:validation} shows that interaction parameters and the variance of environmental noise are accurately estimated when the mean of interactions is non-zero. We have validated our inference approach for a null model with no interactions ($\mu=\sigma_{\rm int}=0$). In that case, the variance of interactions is overestimated (because the slope of the second-stage regression is forced to be positive, SI Appendix, section~S3 and Fig.~S2).
 
Natural communities are open systems whose dynamics are continuously influenced by immigration. As a result, ecological surveys typically include both occasional species, whose presence is largely driven by dispersal, and a persistent set of $S$ species that forms the community ``core''~\cite{Hanski1982,Hanski1982Bumblebees,HanskiGyllenberg1993,MagurranHenderson2003,magurran2007species}. Experimental studies have demonstrated that ecological communities can shift from being predominantly shaped by biotic interactions to being primarily governed by immigration processes as the degree of community isolation decreases. Under such conditions, an established and persistent core of resident species becomes progressively supplemented by an increasing number of transient, incidentally occurring species~\cite{loke2023unveiling}. Because interaction inference is meaningful only for species whose abundances are primarily shaped by interactions, identifying the persistent core is crucial for interaction inference. Consequently, statistical methods must distinguish the interaction-dominated core community from transient species maintained by immigration. The SI Appendix, section~S3.1 (and Figs.~S4--S5), describes a procedure to estimate an interval for the core size $S$. For that subset of \emph{interacting} species, the regression-based methodology is applied to infer interaction parameters.
 
\begin{figure}[t!]
\centering
\includegraphics[width=\columnwidth]{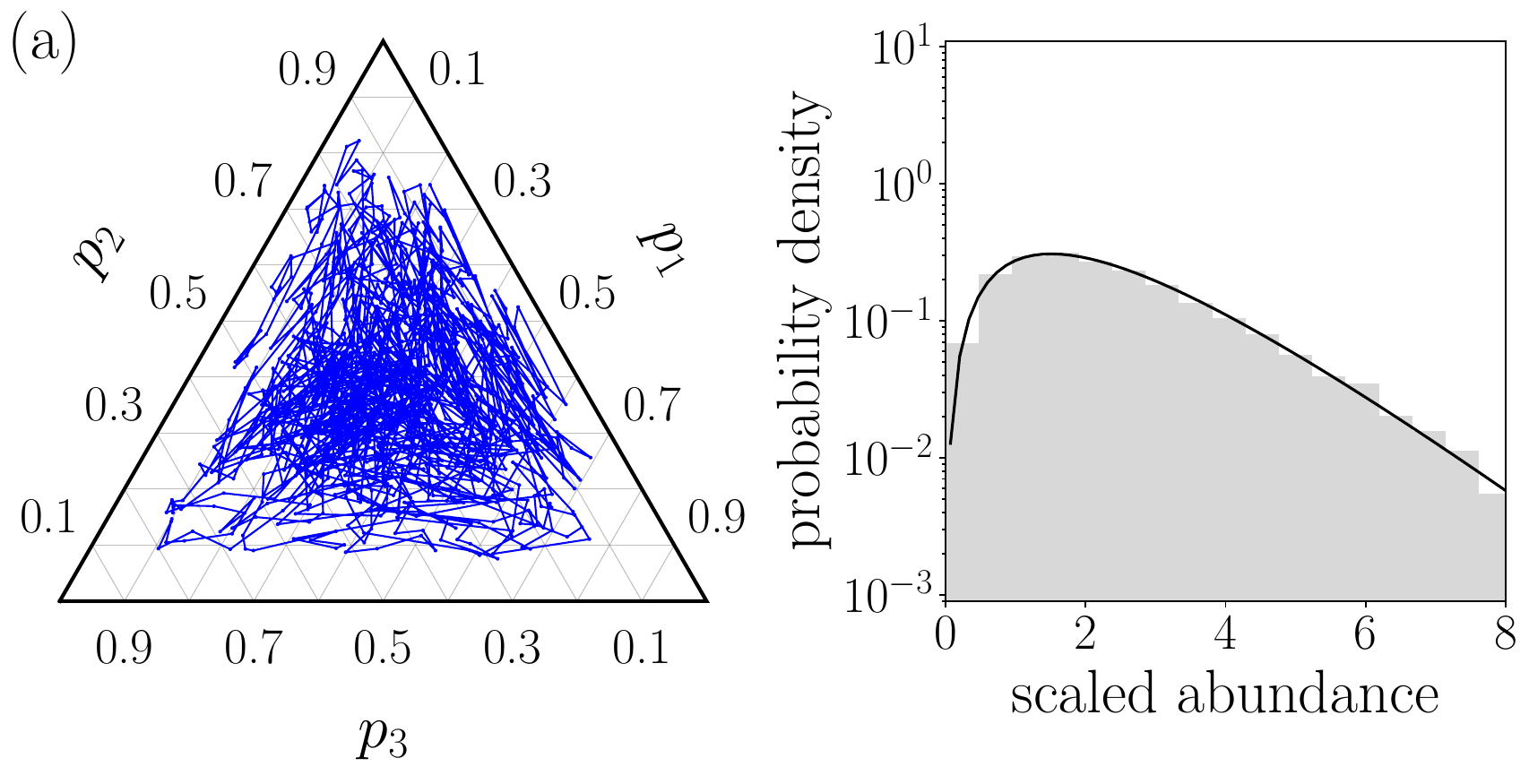}\\
\includegraphics[width=\columnwidth]{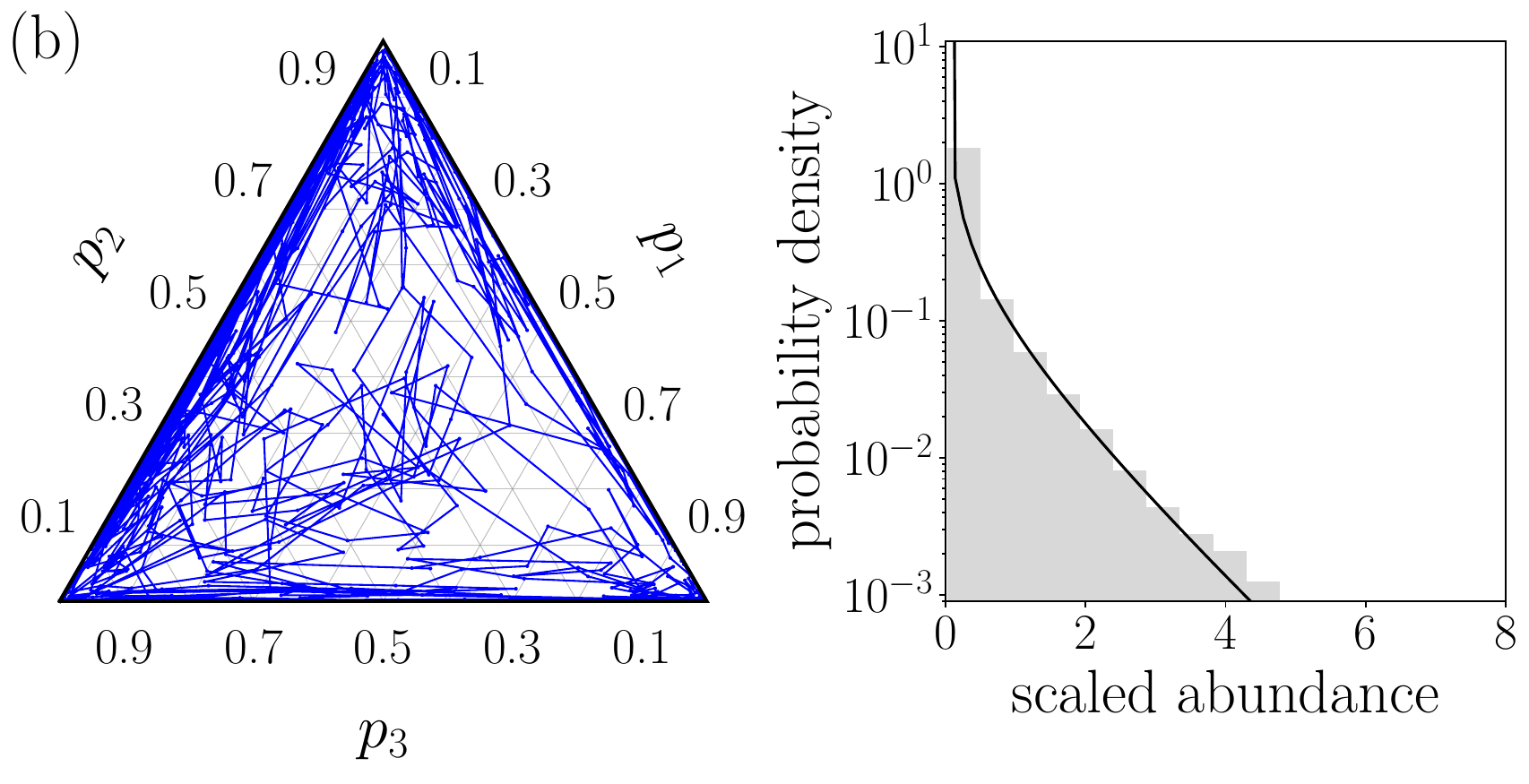}
\caption{\textbf{Ecological phases.} (a) Stable coexistence phase ($\alpha > 1$), with relative abundances fluctuating in the interior of the simplex (here we show three relative abundances of a simulated model trajectory); the SAD in this regime has a mode at strictly positive abundance (right). (b) Quasi-extinction phase ($\alpha < 1$), for which relative abundances are close to zero with high probability; the corresponding SAD peaks at zero abundance instead (right). Abundances are divided by the scale parameter $\theta$ of the Gamma distribution.}
\label{fig:twophases}
\end{figure} 
 
\begin{figure*}[t!]
\centering
\includegraphics[width=0.34\textwidth]{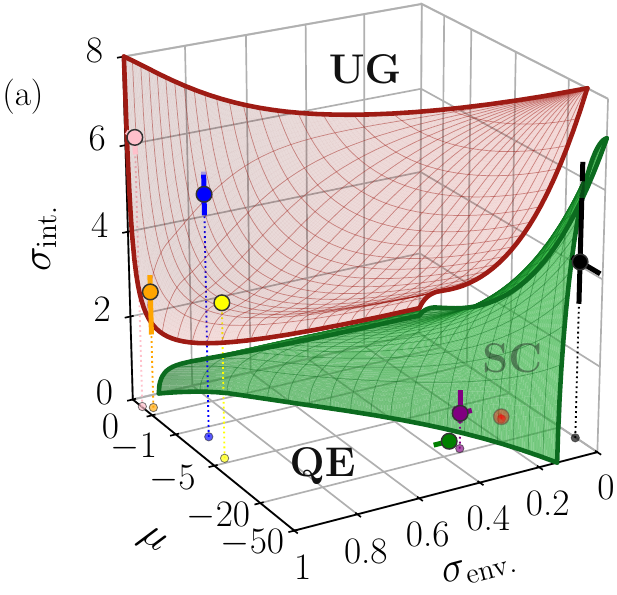}
\raisebox{.88cm}{\includegraphics[width=0.315\textwidth]{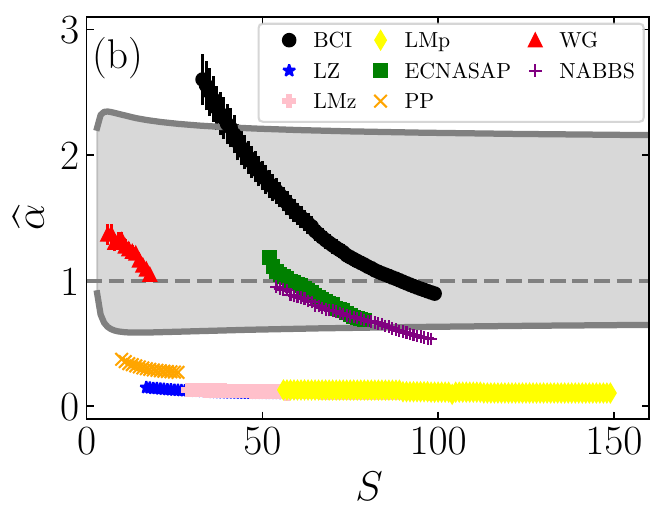}}
\includegraphics[width=0.33\textwidth]{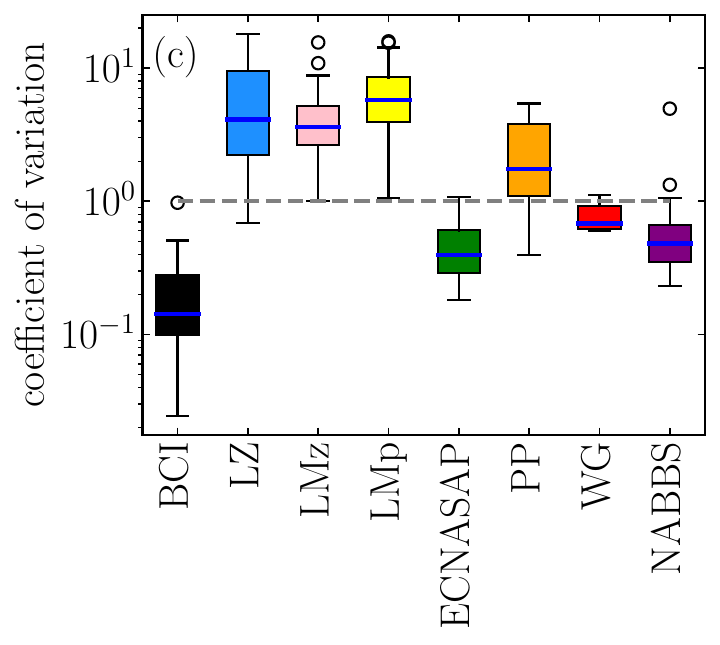}
\caption{\textbf{Phase transitions.} (a) Phase diagram in units of $r=1$ (the $\mu$ axis is represented in logarithmic scale). Three phases arise: stable coexistence (SC, behind the green surface), quasi-extinction (QE, in between of both surfaces), and unbounded growth (UG, above the red surface). Phase transitions (surfaces) are sharp only in the limit $S\to\infty$. Errorbars correspond to the 25\% and 75\% percentiles of the distribution of parameters ($\mu$, $\sigma_{\text{int}}$, $\sigma_{\text{env}}$) for core sizes $S$ compatible with data. The risk of extinction of natural communities is evaluated depending on the phase (SC or QE) they belong to. (b) Estimated Gamma's shape parameter $\widehat{\alpha}$ as function of the core sizes ($S$) consistent with data. The grey area marks, in terms of $\widehat{\alpha}$, the width of the transition between SC and QE phases (observe that the transition width converges to zero very slowly). Communities below the lower grey curve are subject to high extinction risk, whereas species in communities above the upper curve are guaranteed to fully coexist. (c) Boxplots of the coefficient of variation (CV) of species abundances across all datasets. Each boxplot represents the distribution of CV values (in log scale) calculated across time for individual species. The dashed horizontal line represents the value $\text{CV}=1$. Datasets: Barro Colorado Island (BCI); Lake Z\"urich zooplankton (LZ); Lake M\"uggelsee zooplankton (LMz); Lake M\"uggelsee phytoplankton (LMp); Groundfish (ECNASAP); Rodents (PP); Women gut (WG); Bird Survey (NABBS).}
\label{fig:PD}
\end{figure*} 
 
\subsection*{Environmental stochasticity induces ecological phase transitions}
The analysis of the Fokker--Planck equation associated with the SDE~\ref{eq:DMFT} in the stationary regime yields the expected SAD as a Gamma law (SI Appendix, section S4, Eqs.~S24--S25): $P(x) = \frac{x^{\alpha-1}e^{-x/\theta}}{\Gamma(\alpha)\theta^\alpha}$, with parameters $\alpha = \frac{r-\sigma_{\mathrm{env}}^2+\mu M}{\sigma_{\mathrm{env}}^2 + \frac12\sigma_{\mathrm{int}}^2\Sigma}$, and $\theta = \sigma_{\mathrm{env}}^2 + \frac12\sigma_{\mathrm{int}}^2\Sigma$. The expected value of the Gamma law is $M=\alpha\theta$. In terms of model parameters, the first two moments satisfy the self-consistent relations
\begin{equation}\label{eq:MSigma}
M = \frac{r-\sigma_{\mathrm{env}}^2}{1-\mu}, \qquad \text{and}\qquad \Sigma = \frac{2(M+\sigma_{\mathrm{env}}^2)}{2M^{-1}-\sigma_{\mathrm{int}}^2}.
\end{equation}
This form of the distribution is compatible with data, as previously reported~\cite{VolkovEtAl2009,grilli2020macroecological,suweis2024generalized}, see SI Appendix, section~S4.2, Fig.~S8. Note that, for a Gamma distribution, the coefficient of variation of abundance is exactly $\text{CV}=1/\sqrt{\alpha}$. The shape parameter is therefore related to the variability of the representative species' time series in the sense of reference~\cite{LoreauMazancourt2013,Arnoldi2016}, which provides a direct bridge between the phases derived below and the variability-based metrics used to quantify stability from abundance time series.
 
The theory predicts three phases separated by analytical transition boundaries. Observe that species coexistence requires that $\mu < 1$ and $r-\sigma_{\text{env}}^2\ge 0$, so that the mean community abundance $M$ at stationarity remains non-negative---for $\mu > 1$ and $r-\sigma_{\text{env}}^2 < 0$, $M$ is still positive, but in this case there is a bifurcation which stabilizes the extinction equilibrium $x=0$ of the effective dynamics, which leads to the nonphysical extinction of the entire community (SI Appendix, section~S4.1). Additionally, abundance fluctuations must remain finite: the transition from finite to divergent second moment $\Sigma$ determines the transition from bounded to unbounded growth (SI Appendix, section S4.2 and Fig.~S6). Bounded dynamics occurs when fluctuations are well defined, for which the denominator of $\Sigma$ has to be positive: $2M^{-1}-\sigma_{\text{int}}^2>0$. This implies (see Eq.~\ref{eq:MSigma}):
\begin{equation}
r\sigma_{\mathrm{int}}^2 < \frac{ 2(1-\mu) }{ 1 -  \sigma_{\mathrm{env}}^2/r}.
\label{eq:UG}
\end{equation}
Violation of Eq.~\ref{eq:UG} leads to a nonphysical phase characterized by divergent abundances. Because $r\ge \sigma_{\text{env}}^2\ge 0$, we can re-scale variances with $r$ as shown in Fig.~\ref{fig:validation}, and the stability behavior is only determined by three parameter combinations: $\sigma_{\mathrm{env}}^2/r$, $r\sigma_{\mathrm{int}}^2$, and $\mu$. Equivalently, from now on we set $r=1$.

Within the bounded regime, coexistence properties are controlled by the Gamma shape parameter $\alpha$, which fixes the behavior of the SAD at vanishing abundance. As $x\to 0$, the Gamma density diverges for $\alpha<1$ and vanishes for $\alpha>1$. This implies that the mode of the distribution is located at $x=0$ in the former case, and at a strictly positive abundance $x=(\alpha-1)\theta>0$ in the latter. The value $\alpha=1$ therefore separates two qualitatively different situations: for $\alpha>1$, species are repelled from the extinction boundary, because $P(x)\to 0$ as $x\to 0$, whereas for $\alpha<1$ probability accumulates at arbitrarily low abundances, because $P(x)\to \infty$ as $x\to 0$ in that case. We call the first regime the ``stable coexistence'' (SC) phase, and refer to the second one as the ``quasi-extinction'' (QE) phase. The typical dynamics of these two phases is illustrated in Fig.~\ref{fig:twophases}: for $\alpha>1$, relative abundances fluctuate in the interior of the configurational space, with positive abundances, whereas if $\alpha<1$, trajectories repeatedly approach the boundaries of the space, associated to zero abundance. In terms of the variability-based stability metrics (coefficient of variation of the time series), the SC (QE, respectively) phase corresponds to a representative species with coefficient of variation less (greater) than one.

The same threshold is recovered from the probability of feasibility~\cite{servan2018coexistence,lechon2026robust}, defined as the probability that every species in the community remains above a quasi-extinction threshold set at a fraction $1/S$ of the mean abundance $M=\alpha\theta$. In the limit $S\to\infty$, this probability takes the values $p_F=1$ for $\alpha>1$ and $p_F=0$ for $\alpha<1$ (Materials and Methods), so that the whole set of species coexists stably in the SC phase, whereas full coexistence becomes very unlikely in the QE phase, where a number of species have abundances close to zero. This agreement is not automatic, since the threshold must be allowed to shrink as the community grows: $1/S$ is the only such scaling for which the feasibility criterion reproduces the transition at $\alpha=1$ predicted by the SAD (SI Appendix, section S4.3). 

The transition between SC and QE in the limit of high community diversity occurs at $\alpha=1$, yielding the condition
\begin{equation}\label{eq:MAth}
r\sigma_{\text{int}}^2 \le \frac{[1+(\mu-2)\sigma_{\text{env}}^2/r](1-\mu)}{(1-\sigma_{\text{env}}^2/r)^2}.
\end{equation}
The regime shifts determined by Eqs.~\ref{eq:UG} and~\ref{eq:MAth} are represented (for $r=1$) in a phase diagram (Fig.~\ref{fig:PD}a) as function of $\mu$, $\sigma_{\text{int}}$, and $\sigma_{\text{env}}$. The intensity of environmental stochasticity ($\sigma_{\text{env}}$) systematically enlarges the QE region, indicating that environmental variability destabilizes coexistence even when deterministic interactions remain bounded.
 
In the SI Appendix, section~S4.3 and Fig.~S9, we show that the transition from SC to QE is not sharp for finite $S$ unless the species richness takes very large values. The values of $\alpha$ such that the probability of feasibility of the whole community $p_F$ is below $1/S$ are compatible with close-to-zero abundances (because, on average, one species is expected to survive), while the values of $\alpha$ such that $p_F>1-1/S$ are compatible with a full coexistence regime. In between, the broadness of the transition between the two phases is encapsulated (Fig.~\ref{fig:PD}b). 

\subsection*{Application to natural ecological communities}
 
We applied the methodology to 8 datasets, comprising a variety of habitats and taxa. Previous work has reported transitions between alternative states in the following cases: the zooplankton community in Lake Z\"urich (LZ), two communities in Lake M\"uggelsee (LMp for the phytoplankton and LMz for the zooplankton subcommunities), the Women Gut microbiota dataset (WG) and the rodent abundance dataset (PP). In contrast, the tropical forest in Barro Colorado Island (BCI) has been identified as an equilibrium community under low disturbance (see Materials and Methods for dataset descriptions and references). The remaining time-series data were used for stability prediction: we analyzed  abundance data from the North American Breeding Bird Survey (NABBS), and time series of groundfish abundance (ECNASAP). Datasets were selected according to three criteria: (i) sufficiently dense temporal or spatial sampling for the analyses, (ii) previous ecological information relevant to community stability, and (iii) representation of a broad range of taxa and ecosystems. 

Core sizes compatible with abundance data vary within an interval, leading to a distribution of parameter estimates, shown as errorbars in the phase diagram for each dataset (Fig.~\ref{fig:PD}a). The median of the distribution (marked with a dot) locates each community in the diagram. Whereas BCI, WG, ECNASAP and NABBS mostly lie within the broad boundary of the SC--QE transition, the remaining communities clearly fall in the QE phase. 

Colored lines in Fig.~\ref{fig:PD}b show, for each natural community, maximum likelihood (ML) estimates of the shape parameter $\alpha$ as a function of core size $S$, for all core sizes leading to consistent parameter estimates (SI Appendix, section~S3.1). The estimated values of $\alpha$ are compatible with the empirical distribution of variability measures (coefficient of variation of each species' time series) obtained directly from data (Fig.~\ref{fig:PD}c): datasets whose coefficient of variation is distributed well above $1$ (below $1$) are compatible with estimates $\widehat{\alpha}<1$ ($\widehat{\alpha}>1$), see also SI Appendix, section S4.2. 
 
Results of ML fitting of the Gamma distribution to empirical SADs, constructed as histograms formed by pooling abundances across all time steps, are reported in SI Appendix, section S4.2 (Fig.~S8). The existence of a mode at positive values is compatible with the classification of communities according to their stability regime in the phase diagram. 

Estimated interquartile intervals (compatible with the interval for core sizes, see Fig.~\ref{fig:PD}b and SI Appendix, section S3.1) for $\mu$, $\sigma_{\text{int}}$, and $\sigma_{\text{env}}$ are reported in Table~\ref{tab:risk}, together with an interval for the estimated shape parameter $\alpha$. The single-species extinction risk is quantified as $1-p_1$, $p_1$ being the probability that the effective species has feasible abundance (Materials and Methods). To be more precise, $1-p_1$ is the stationary probability that the abundance of the representative species falls below a fraction $1/S$ of the community mean, and should be interpreted as a quasi-extinction risk rather than as a probability of extinction. As Fig.~\ref{fig:PD}b anticipates, the BCI community has the lowest extinction risk, a consistent result given the slow demographic turnover of tropical forest trees. Aside from BCI, WG, ECNASAP and NABBS, the remaining communities are estimated to have high extinction risk, especially the zooplankton communities, with extinction risks above 0.5 for the effective species.
 
\section*{Discussion}
 
Our results establish a direct connection between abundance fluctuations, interaction statistics, and the stability of species-rich ecological communities. By combining DMFT with statistical inference, we show that the stability of large ecosystems can be characterized without reconstructing the entire interaction network. Instead, a small set of effective interaction statistics inferred from abundance time series is sufficient to predict the dynamical regime of the community and to quantify its expected extinction risk. Remarkably, the fitting procedure also returns the community characteristic time scale $\Delta t$ measured in units of the timescale of self-regulation, a dimensionless quantity that indicates whether a monitoring scheme resolves community dynamics at all~\cite{ontiveros2021characteristic}. 
 
\setlength{\tabcolsep}{2pt}
 
\begin{table*}[ht]
\scriptsize{
    \centering
    \begin{tabular}{c|c|c|c|c|c|c|c|c}
       & BCI & LZ & LMz & LMp & ECNASAP & PP & WG & NABBS \\
       \hline\hline
       $\widehat{\mu}$ & $(-47.72, -24.26)$ & $(-1.74, -1.63)$ & $(0.20, 0.78)$ & $(-4.33, -4.11)$ & $(-14.32, -11.73)$ & $(0.11, 0.40)$ & $(-11.63, -9.64)$ & $(-21.23, -14.79)$ \\
       $\widehat{\sigma}_{\rm int}$ & $(3.14, 6.36)$ & $(5.14, 6.12)$ & $(5.92, 6.49)$ & $(3.45, 3.67)$ & $\sim 0$ & $(1.75, 3.16)$ & $(3.38, 8.72)\times 10^{-2}$ &  $(0.62,1.38)$ \\       
       $\widehat{\sigma}_{\rm env}$ & 
       $(1.21, 2.65)\times 10^{-3}$ & 
       $(0.91, 0.93)$ & $(0.997, 0.999)$ & $(0.951, 0.953)$ & $(0.31, 0.38)$ & $(0.97, 0.98)$ & $(0.116, 0.121)$ & $(0.29, 0.35)$ \\ 
       $\widehat{\alpha}$ & $(1.07,1.85)$ & $(0.10,0.12)$ & $(0.11, 0.12)$ & $(0.11,0.13)$ & $(0.77,0.99)$ & $(0.28, 0.34)$ & $(1.16, 1.32)$ & $(0.62, 0.80)$ \\
       $1-p_1$ &  $(1.17,8.76)\times10^{-3}$ & $(0.54, 0.55)$ & $(0.510, 0.516)$ & $(0.47, 0.49)$ & $(1.69, 3.22)\times 10^{-2}$ &   $(0.315,0.319)$ & $(4.07, 4.83)\times10^{-2}$ & $(3.05, 5.10)\times10^{-2}$\\\hline\hline
    \end{tabular}}
    \caption{Interquartile intervals for empirical community fitted parameters (in units of $r=1$) and estimated species extinction risk (last row).}
    \label{tab:risk}
\end{table*}
 
Our results demonstrate that environmental variability increases the probability that species spend prolonged periods at very low abundances. This is apparent in the phase diagram (Fig.~\ref{fig:PD}a): as long as $\sigma_{\text{env}}$ increases, the region associated to SC shrinks. This result provides a framework to quantify the role of environmental variability in shaping extinction risk and community persistence from time series of species abundances. 

\subsection*{Population dynamics}

Our framework also occupies a specific position within the stochastic GLV literature it builds upon. The stochastic logistic model~\cite{grilli2020macroecological} reproduces the single-species macroecological laws with environmental noise and self-regulation alone, establishing that they do not require inter-specific interactions, but for that same reason it cannot address coexistence, abundance correlations, or the interaction statistics that govern them~\cite{CamachoMateu2024}. Conversely, quenched treatments of random GLV dynamics~\cite{Bunin2017,galla2018dynamically} characterize the phases generated by interactions and their stability while omitting environmental variability. Suweis et al.~\cite{suweis2024generalized} endow the interactions themselves with a finite correlation time, of which Eq.~\ref{eq:SDE_GLV} is the vanishing-correlation-time limit. In their treatment, however, environmental stochasticity is not included explicitly, whereas here it acts as an independent source of variability on the per-capita growth rate, making $\sigma_{\text{env}}$ and $\sigma_{\text{int}}$ separately identifiable. These approaches are generative, asking which ingredients reproduce observed patterns. We instead use the same class of models as an inference tool, estimating the parameters compatible with abundance time series, and then use the estimates to predict stability. 

\subsection*{Alternative approaches to interaction inference}
 
Although we have already discussed the complexity of interaction inference from abundance data, it is convenient to put our framework within the context of recent research. In reference~\cite{CamachoMateu2024}, the authors developed a Bayesian methodology that uses pairwise correlation data from microbial communities to estimate ensembles of interaction matrices. As in our approach, it is based on a stochastic Lotka--Volterra equation with environmental noise uncorrelated across species, and it renounces identifying individual pairwise couplings, sampling instead an \emph{ensemble} of matrices compatible with the data. It differs with our approach in five respects. First, their framework treats empirical samples as independent realizations, whereas ours uses the temporal ordering explicitly and, as a by-product, recovers the typical timescale of the dynamics relative to self-regulation~\cite{ontiveros2021characteristic}. Second, our reduction shows that macroscopic behavior depends on the interaction matrix only through $\mu$ and $\sigma_{\text{int}}$, estimated without sampling full matrices. Third, in~\cite{CamachoMateu2024} the environmental noise intensity is fixed \emph{a priori} and only the interaction matrix is inferred, whereas we estimate the interaction statistics and the environmental variability $\sigma_{\text{env}}$ jointly. Fourth, we treat interactions in an annealed regime, as in the uncorrelated limit of~\cite{suweis2024generalized}, whereas in~\cite{CamachoMateu2024} a quenched approach is used. Fifth, and most importantly, their prior vanishes for matrices that do not yield a feasible, stable equilibrium, so the inferred ensembles are stable by construction and cannot be used to determine community stability. In ours, the stability phase is an output of the inference, depending only on the estimates of community parameters, which is what allows us to attach a quantitative extinction risk to each community.

The main empirical conclusion in~\cite{CamachoMateu2024} is that the networks compatible with observed microbial correlations are sparse, with an effective connectance of a few percent. At the mean-field level, however, a sparse matrix in which a fraction $C$ of the entries are drawn with standard deviation $\sigma$ is indistinguishable from a dense matrix with standard deviation $\sigma\sqrt{C}$, since only $S\,\text{Var}[a_{ij}]=SC\sigma^2\equiv\sigma_{\text{int}}^2$ enters the effective dynamics. Sparsity and the variability of interaction strengths are therefore not separately identifiable from abundance patterns, and $\sigma_{\text{int}}$ must be read as the only compound quantity that can be jointly determined. The same logic underlies our treatment of the community core: sparsity reduces the number of couplings each species feels, whereas restricting the analysis to the persistent core reduces the number of species whose dynamics are interaction-dominated at all, the remainder being sustained by immigration~\cite{loke2023unveiling}. Both approaches identify the effective number of interacting species that abundance data can support. They are thus complementary: reconstructing entire matrices gives access to the sign structure or modularity of the network, which our reduction cannot address, whereas ours affords parameter identifiability, closed-form uncertainty quantification, estimation of the environmental noise intensity, and a falsifiable prediction of the stability regime.

\subsection*{Immigration- versus interaction-driven communities}

The communities we analyze are open, and the species they contain are not all shaped by the same processes: alongside a persistent, interaction-dominated core, ecological surveys record occasional species whose presence is largely set by dispersal from the regional pool~\cite{Hanski1982,MagurranHenderson2003,loke2023unveiling}. Restricting the inference to the core is therefore not merely a technicality of our inference framework; the core size $S$ is itself an ecological quantity, measuring how many species have abundances governed by biotic interactions rather than by immigration. Enlarging the core beyond its interaction-dominated part progressively adds species whose fluctuations are dispersal-generated; since immigration is not modeled explicitly, that variability is absorbed by $\sigma_{\text{env}}$ (which thereby increases), displacing the estimated community towards the QE region of parameter space. This is apparent also from Fig.~\ref{fig:PD}b, as the core size $S$ augments. Previous approaches mostly avoid the determination of the subset of interacting species, adjusting effective models to the whole set of abundances. Importantly, our approach provides a systematic methodology to extract the core of interacting species from time series.

\subsection*{Alignment of the stability predictions with previous work}
 
The application to empirical communities illustrates both the strengths and limitations of our approach based on estimating the distribution of interactions, not pairwise interactions themselves. Although the datasets analyzed span ecosystems with contrasting taxonomic composition and temporal resolution, the inferred parameters consistently locate each community within a stability phase. For datasets with independent evidence concerning their stability, the regimes inferred by our framework generally agree with previous empirical assessments, with WG as a possible exception (see below). For BCI, previous work reported abundance fluctuations around stationary mean values and a community close to equilibrium under currently low disturbance levels~\citep{Cavagna2023,Ruger2020}, consistent with our classification as stable coexistence with low environmental variability. For PP, LZ, LMz, and LMp, multistability, regime shifts, or persistent seasonal differences in composition have been reported~\citep{Korner2001,Christensen2018,Anneville2004}; although not equivalent, each involves recurrent differences among community configurations, broadly consistent with the QE regime. The net positive biotic dependencies inferred for PP and LMz (see estimates for $\mu$, Table~\ref{tab:risk}) are particularly noteworthy. Because the estimated interaction parameter represents an aggregate community-level effect, these results do not necessarily imply direct pairwise facilitation (in addition, the estimated deviations $\sigma_{\text{int}}$ are large in these two cases, also compatible with negative interactions). In PP, climatic variability, resource pulses, and compensatory increases by functionally similar species could generate positive net effects despite direct competition \cite{Ernest2000,Brown2002,Goheen2005}. Similar effects could arise in LMz through shared environmental responses, nutrient recycling, or trophic indirect interactions. The remaining communities provide negative estimates of $\mu$, with small or moderate dispersion of interaction strengths, thus being compatible with dominant competitive interactions.

WG presents a less straightforward case: because enterotype changes occur over time, we expected a QE regime, yet its inferred dynamics are closer to stable coexistence. This would be consistent with enterotypes not representing alternative stable states, since long-term transitions between persistent enterotype states have not been observed~\citep{Vandeputte2021}. Notable in particular is the identification of communities that have species with potential extinction risks. Whether these inferred regimes anticipate future community reorganization remains an important question for upcoming studies. Following our framework, future studies may assess stability regimes at large scale using available time series from a variety of habitats comprising different taxa \cite{Dornelas2025}.

\subsection*{Limitations and extensions}

Our description retains only the first two moments of the distribution of couplings, so that trophic hierarchies, nested architectures, modularity, heavy-tailed coefficients, or any other kind of interaction structure are invisible to it. This is a genuine limitation for vertically structured food webs, in which predator--prey pairs of opposite sign dominate the spectrum~\cite{AllesinaTang2012}, and less severe for the horizontal, competition-dominated assemblages most of our datasets represent~\cite{Bunin2017}, for which most estimates of $\mu$ are indeed negative. Machine learning approaches may help extend our methodology to assess the effect of network structures on stability. 

With regard to how fluctuations are treated, environmental noise is taken uncorrelated across species and white in time, whereas real forcing is seasonal and shared among related taxa~\cite{SireciEtAl2023}. Thus, an important refinement of the framework would consider correlated noise. An additional simplification is the absence of demographic stochasticity, which scales as the square root of abundance and therefore dominates precisely in the low-abundance regime that defines the quasi-extinction phase. Including it would presumably shift the SC--QE phase boundary. 

Two extensions appear especially valuable. First, the core--satellite decomposition we adopt draws a sharp line between interaction-dominated and dispersal-maintained species, whereas the transition between the two regimes is gradual and depends on isolation, on environmental variability and on the variance of interactions~\cite{loke2023unveiling}. Adding an explicit immigration term to the effective dynamics would make this dependence quantitative, would regularize the boundary at zero abundance, and would remove the need to treat the core size as a nuisance parameter. It would also sharpen the interpretation of $\sigma_{\text{env}}$, which currently absorbs any unmodeled source of variability, including growth-rate heterogeneity and observation error. More importantly, inferring immigration would allow to separate communities mostly driven by ecological drift, from those that are partially or totally determined by interactions. Second, observation error and irregular census intervals are pervasive in monitoring data and bias regression estimates of density dependence towards spurious regulation~\cite{KnapeDeValpine2012}, so that separating process noise from observation error, as in state-space formulations of community time-series models~\cite{IvesEtAl2003}, is an important extension. 

Because the method requires only abundance time series and returns a small number of interpretable parameters, it can be applied systematically to the thousands of assemblages compiled in biodiversity databases~\cite{Dornelas2025}, and to controlled microcosm experiments in which the predicted phases have begun to be mapped directly~\cite{HuEtAl2022}.
 
Rather than viewing stability as a property that can only be assessed through detailed knowledge of ecological interactions, our results provide solid theoretical support for inferring stability regimes directly from community dynamics, a characterization that agrees with the variability-based view of stability~\cite{LoreauMazancourt2013,Wang2017,Arnoldi2016}. As long-term biodiversity monitoring becomes increasingly widespread, our contribution offers a simple framework for identifying communities approaching instability, quantifying extinction risk, and evaluating the consequences of environmental change for ecosystem persistence.
 
\section{Materials and methods}
\subsection*{Multi-species GLV dynamics}
Before abundance re-scaling, the $S$-species GLV dynamics with environmental noise is defined as
\begin{align}\label{eq:SDE_GLV1}
	&\dot{q}_i(t) = q_i(t)\left(r_i+\sum_{j=1}^S A_{ij}(t)q_j(t) + \sqrt{2\sigma_{\rm env}^2}\xi_i(t)\right), \,\, i=1,\dots, S,\nonumber\\ 
    &\langle \xi_i(t)\rangle = 0,\quad \langle \xi_i(t)\xi_j(t^{\prime})\rangle = \delta_{ij}\delta(t-t^{\prime}), 
\end{align}
where the second-row conditions imply that the environmental noise has mean zero and is uncorrelated both at different times and among species (here, $\delta_{ij}$ and $\delta(x)$ stand for the Kronecker and Dirac deltas, respectively). Here, $A_{ii}(t)<0$ to quantify intraspecific interactions. Upon rescaling abundances as $x_i(t)=-A_{ii}(t)q_i(t)$, and letting $a_{ij}(t):=-A_{ij}(t)/A_{ii}(t)$, then Eq.~\ref{eq:SDE_GLV1} reduces to Eq.~\ref{eq:SDE_GLV}. Notice that each coefficient $a_{ij}(t)$ preserves the sign of $A_{ij}(t)$ since all the elements of the diagonal in the original interaction matrix are negative. Because $r_i$ are Gaussian variables with mean $r$ and an unspecified variance, one can add this source of variability to the environmental noise term, $\sqrt{2\sigma_{\rm env}^2}\xi_i(t)$. This addition would effectively modify the intensity of the environmental noise, so we consider that the variability in growth rates is subsumed in the variance of environmental fluctuations. Without loss of generality, we assume that the intrinsic growth rate is equal to $r$ across species.
 
Pairwise interactions $a_{ij}(t)$ are time-varying, delta-correlated random variables, forming a time-dependent random matrix with off-diagonal correlation $\rho$. Thus, the $z_{ij}(t)$ in Eq.~\ref{eq:alpha} satisfy:
\begin{equation}
\langle z_{ij}(t)z_{k\ell}(t^{\prime})\rangle = \left(\delta_{ik}\delta_{j\ell} +\rho\delta_{i\ell}\delta_{jk}\right)\delta(t-t^{\prime}). 
\end{equation}
Eq.~\ref{eq:alpha} implies that the expectation of interactions is ${\mathbb E}[a_{ij}]=\mu/S$, and the variance of the distribution is $\text{Var}[a_{ij}]=\sigma_{\rm int}^2/S$. The scalings of the mean and variance of $a_{ij}$ with $S$ are standard in random matrix theory, and allow to evaluate properly the high diversity limit, $S\to\infty$.
 
The derivation of the effective species dynamics (cf. Eq.~\ref{eq:DMFT}, and its generalization for non-zero values of the correlation $\rho$), using DMFT, is left to the SI Appendix, section~S2.
 
\subsection*{Probability of feasibility}

We consider the effective species to be quasi-extinct when its abundance falls below a fraction $\epsilon$ of the community mean, i.e.\ below $\epsilon M$, with $M=\alpha\theta$ the mean abundance at stationarity. The stationary probability that the effective abundance lies above that threshold, according to the Gamma distribution, is $p_1=\Pr\!\left\{x\ge \epsilon M\right\}=\Gamma(\alpha,\epsilon\alpha)/\Gamma(\alpha)$, where $\Gamma(\alpha,z)$ is the upper incomplete Gamma function and $\Gamma(\alpha)$ the complete one. Let $N_{<}$ be the expected number of quasi-extinct species. By linearity of expectation, in the DMFT framework it holds that $\langle N_{<}\rangle=S\,(1-p_1)$. The probability of feasibility of the community is $p_F=p_1^S$. We are interested in evaluating the extinction risk $1-p_1$ and $\langle N_{<}\rangle$ when $\epsilon\to 0$ and $S\to\infty$. The threshold $\epsilon$ must be
allowed to shrink as the community size $S$ grows, so that the natural scaling of the threshold (relative to the mean abundance $M$) allows to determine if a species will be close to extinction.

In the SI Appendix (section~S4.3), we show that $\epsilon = 1/S$ is the natural resolution scale of the threshold for a community of $S$ members: it scales with richness in the same way as the per-species share of a fixed total abundance does. That particular scaling is immediately recovered because the transition between SC and QE phases at $\alpha=1$ is only obtained if $\epsilon=1/S$.  Accordingly, in the limit $S\to\infty$, the expected number of species below that threshold, $\langle N_{<}\rangle$, vanishes for $\alpha>1$, diverges for $\alpha<1$, and equals exactly one at $\alpha=1$, as expected for the SC and QE phases, respectively. The SC--QE large-$S$ boundary $\alpha=1$ is therefore realized when one single species of the community is expected to be quasi-extinct. The extinction risks reported in Table~\ref{tab:risk} are thus obtained as $1-p_1=1-\Gamma(\alpha,\alpha/S)/\Gamma(\alpha)$.

\subsection*{Data, Materials, and Software Availability}
We analyzed species-abundance data from natural communities sampled repeatedly through time or across space. The Women Gut (WG) dataset comprises longitudinal gut-microbiota observations from participant 820 in the study by Vandeputte et al.~\cite{Vandeputte2021}. This participant was sampled over approximately two months and exhibited several enterotype transitions. Previous work has discussed such transitions in relation to alternative microbiome states~\cite{Gonze2017}. The Lake M\"uggelsee dataset comprise phytoplankton (LMp) and zooplankton (LMz) records, sampled over more than 20 years~\cite{Langenhaun2020}. Previous analyses of this system reported evidence consistent with alternative community states~\cite{Korner2001}. The Lake Z\"urich (LZ) dataset described by Pomati et al.~\cite{Pomati2012,Pomati2015} comprises approximately 35 years of zooplankton observations. Previous work identified seasonally differentiated community configurations in this system~\cite{Anneville2004}. We also analyzed records from the North American Breeding Bird Survey (NABBS)~\cite{Ziolkowski2025}. For this analysis, we restricted the data to survey locations in West Virginia belonging to a single ecoregion. The rodent dataset (PP) assembled by Ernest et al.~\cite{Ernest332783} spans nearly 50 years of community observations; previous work has associated changes in this community with multistability~\cite{Christensen2018}. The Barro Colorado Island 50-ha forest dynamics plot (BCI) provides spatially resolved tropical-tree census data collected over approximately 35 years~\cite{Condit2019}. We treated each whole-plot census as a single temporal observation and did not use spatial subdivisions as replicates. Finally, we analyzed a groundfish-abundance dataset (ECNASAP) spanning nearly 25 years~\cite{Brown2005}.

The code to reproduce all the results reported is provided in the repository \url{https://github.com/vjontiveros/annealed-interactions/}.\par

\section{Auhtor contributions}
A.M. and J.A.C. designed research. A.M. and J.A.C. performed analytical derivations and designed the statistical framework. A.M. drafted the code for simulation and statistical analyses. V.J.O. acquired and prepared the data for the statistical analysis. J.A.C. drafted the manuscript. A.M., V.J.O., D.A., and J.A.C. contributed to the mathematical derivations, the implementation of the code, edited the manuscript, and discussed and interpreted the results.

\section{Acknowledgements}
This work has been supported by grants PRIORITY (PID2021-127202NB-C22), awarded to JAC, and UNIQUE (PID2021-127202NB-C21), awarded to DA, funded by MCIN/AEI/10.13039/501100011033 and ``ERDF. A way of making Europe''.

\bibliography{references_rev}

\includepdf[pages={{},1-15}]{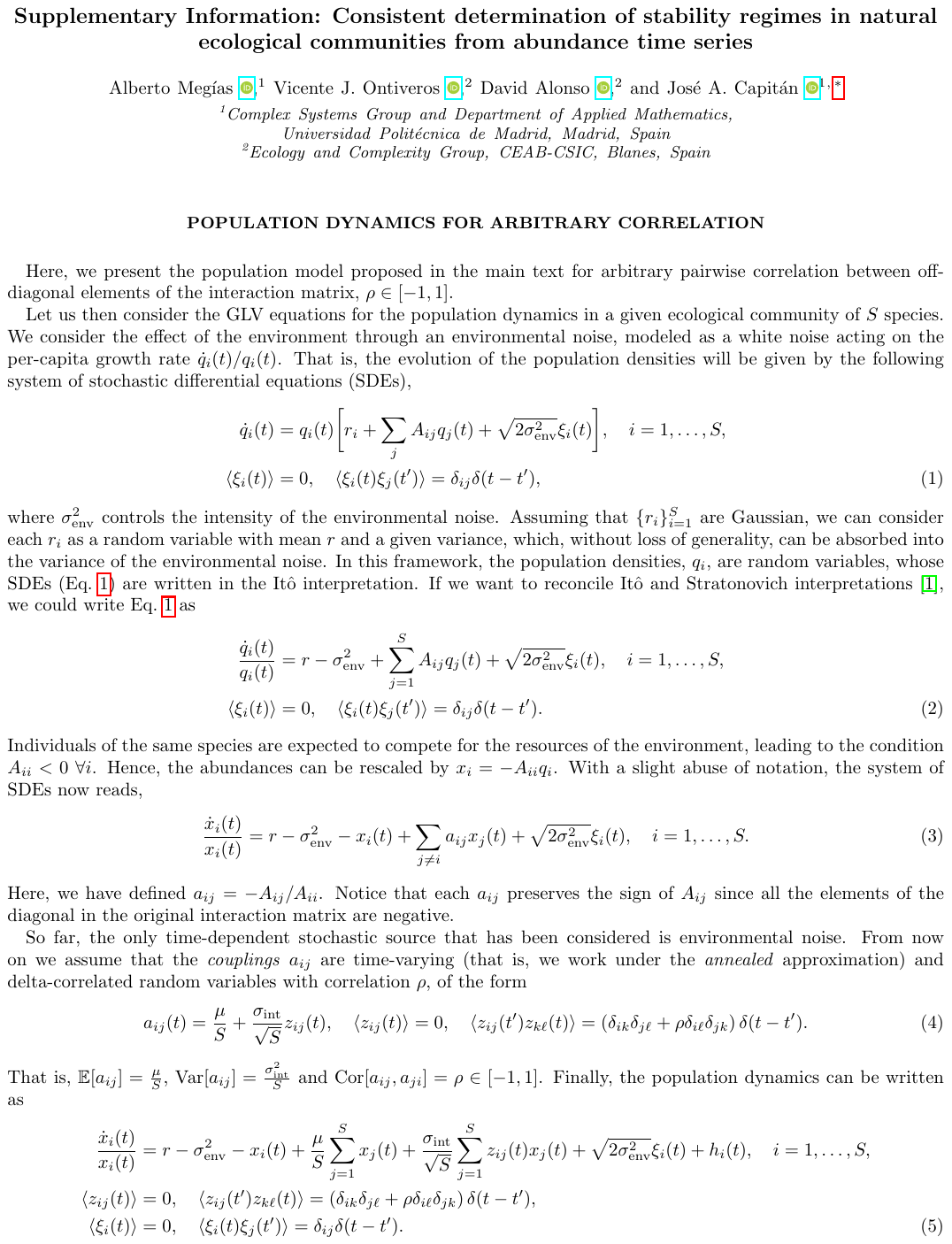} 
\end{document}